\documentclass[aps,prb,twocolumn,showpacs]{revtex4}
\usepackage{amsmath}
\usepackage{dcolumn}
\usepackage{amssymb}
\usepackage{amsfonts}
\usepackage{txfonts}
\usepackage{latexsym}
\usepackage{euscript}
\usepackage[dvips]{graphicx}
\usepackage[T1]{fontenc}
\usepackage{textcomp}

\begin{document}
\title{Frustrated Bose condensates in optical lattices} \author{T. \DJ
uri\'c and D.K.K. Lee} \affiliation{Blackett Laboratory, Imperial
College London, Prince Consort Road, London SW7 2AZ, United Kingdom}
\date{\today}
\begin{abstract}

We study the Bose-condensed ground states of bosons in a
two-dimensional optical lattice in the presence of frustration
due to an effective vector potential, for example, due to lattice
rotation. We use a mapping to a large-$S$ frustrated magnet to study
quantum fluctuations in the condensed state. Quantum effects are
introduced by considering a $1/S$ expansion around the classical
ground state. The large-$S$ regime should be relevant to systems with
many particles per site.  As the system approaches the Mott insulating state, the hole density
becomes small. Our large-$S$ results show that, even when the system
is very dilute, the holes remain a (partially) condensed
system. Moreover, the superfluid density is comparable to the
condensate density. In other words, the large-$S$ regime does not
display an instability to noncondensed phases.  However, for cases
with fewer than 1/3 flux quantum per lattice plaquette, we find that
the fractional condensate depletion increases as the system approaches the Mott
phase, giving rise to the possibility of a noncondensed state before
the Mott phase is reached for systems with smaller $S$.
\end{abstract} 
\maketitle

\section{Introduction}
\label{sec:intro}

Bosonic atoms in optical lattices can display superfluid and Mott
insulating phases.  If the system is rotated, then, in the corotating
frame, this is equivalent to introducing an effective magnetic field
proportional to the rotation frequency \cite{CooperReview, Bhat}.
This is not the only means to introduce a vector potential to a system
of neutral atoms.  This can also be achieved\cite{Furtado,Pachos, Pachos2, Kay} through the interaction
of atomic electric and magnetic moments with an external
electromagnetic field 
(Aharonov-Casher and differential Aharonov-Bohm effects). For atoms
trapped in an optical lattice in two distinct internal states, a
scheme \cite{Jaksch} using two additional Raman lasers combined with
the lattice acceleration or inhomogeneous static electric field has
also been proposed.

Bosonic atoms in an optical lattice can be modeled by a Bose-Hubbard
model. A vector potential introduces an Aharonov-Bohm phase for the
boson hopping from site to site. The wave function is ``frustrated'' if
the phase twists around each plaquette add up to $2\pi\alpha$ for some
non-integer $\alpha$. For a Bose condensate at a low effective
magnetic field, this introduces vortices into the condensate. The
presence of the optical lattice \cite{Sorensen, Bhat} interferes with
the formation of an Abrikosov vortex lattice \cite{Williams,
CooperReview} and quantum fluctuations may be enhanced.  Further, if
the number of vortices becomes comparable to the number of bosons, the
system may enter into a fractional quantum Hall state
\cite{CooperReview, Wilkin, Cooper, Rezayi, Sorensen, Bhat}.  However,
this requires a very high rotation frequency or a low atomic density
which is hard to achieve experimentally.

In this work, we will focus on the experimentally accessible regime
where a condensate still exists to examine whether there are any
precursors to such states in a frustrated Bose condensate.  We study a
two-dimensional (2D) Bose-Hubbard model on a square lattice for a range of
incommensurate filling. In the regime of strong on-site interaction,
the model is analogous to a quantum easy-plane ferromagnet and the
frustration encourages spin twists, i.e., the formation of
vortices in the ground state. We find the classical ground states
using Monte Carlo methods and then we study the quantum fluctuations
around the classical state.  In other words, we work under the
assumption that quantum effects do not change qualitatively the nature
of the ordering obtained for the classical ground
states. Mathematically, this means that we will work in a large-$S$
generalization of the spin model and perform an expansion in $1/S$ to
obtain the quantum effects.  Although our original model corresponds
to small $S$, the large-$S$ approach can be justified if the
perturbative series in $1/S$ converges\cite{Canali, Igarashi,
Chubukov, Kollar,Bernardet}.
In those cases, a spin wave calculation may give
accurate results.

We will study how quantum fluctuations affect the order parameter,
off-diagonal long-range order (ODLRO) and the superfluid fraction for
different degrees of frustration for the whole range of incommensurate
filling. In the spin analog, the incommensurate filling corresponds
to a range of Zeeman field $h$ up to some frustration-dependent
critical field $h_c(\alpha)$. Our calculations were made for
$\alpha=0,1/4,1/3$, and 1/2.

Our results show that the degree of Bose condensation decreases as $h$
increases toward $h_c$. However, it does not vanish at the limit of
$h=h_c(\alpha)$.  This applies to several quantities that we have
calculated: the reduction in the order parameter, the reduction in the
largest eigenvalue of the density matrix, and the sum of the
non-macroscopic eigenvalues of the density matrix.  We also find
similar conclusions for the superfluid fraction --- frustration reduces
the superfluid fraction in the comparison with the unfrustrated case
but there is no vanishing of the superfluid fraction at any $h\leq
h_c$.

The paper is organized as follows. We will outline the model and the
mapping to the quantum spin model in Sec. \ref{sec:model}. We
describe the classical ground states ($S\rightarrow\infty$) of the
spin analog in Sec. \ref{sec:classical}. We introduce the
excitations above the ground state in a $1/S$ expansion in Sec.
\ref{sec:quantum}.  In Secs. \ref{sec:densitymatrix} and
\ref{sec:superfluiddensity}, we calculate the degree of condensation
and superfluidity in the system. We make conclusions about our study
in the final section.
\section{Model Hamiltonian}
\label{sec:model}
For atoms trapped in a two-dimensional optical lattice, we can focus on a
single-band lattice model if the tunneling $t$ between
wells within the lattice is weak compared to the level spacings in
each well. If the tunneling is also weak compared to the repulsive
energy $U$ for two atoms in one well, then strongly correlated ground
states, such as the Mott insulator, appear as well as a superfluid
state.

Many different methods have been proposed to introduce frustration in
the atomic motion. This can be done through rotating the
system\cite{CooperReview} or through the interaction of the atoms with
an external electromagnetic field \cite{Furtado,Pachos, Pachos2, Kay}.
If there is only one species of bosonic atoms, then the system is
described by a Bose-Hubbard model on a square lattice with a complex
hopping matrix element: $H_{\rm Hubbard}=H^{(0)}+V$ with 
\begin{eqnarray}\label{eq:BoseHubbard}
H^{(0)}&=&\frac{U}{2}\sum_i
\hat{a}^{\dagger}_i\hat{a}_i(\hat{a}^{\dagger}_i\hat{a}_i -1) - \sum_i
\mu\hat{a}^{\dagger}_i\hat{a}_i,\nonumber\\ T&=&-t\sum_{\langle
ij\rangle}\left(e^{i\phi_{ij}}\hat{a}_j^{\dagger}\hat{a}_i+\mbox{H.c.}\right),
\end{eqnarray}
where $\mu$ is the chemical potential and $\langle ij\rangle$ denotes
nearest-neighbor sites $i$ and $j$.  The complex tunneling couplings
appear in the Hubbard Hamiltonian due to the presence of the effective
vector potential $\vec{A}$.  When an atom moves from a lattice site at
$\vec{R}_i$ to a neighboring site at $\vec{R}_j$, it will gain an
Aharonov-Bohm phase
\begin{equation}\label{eq:ABphase}
\phi_{ij} =\int_{\vec{R}_i}^{\vec{R}_j}\vec{A}\cdot d\vec{r},
\end{equation}
For neutral atoms with electric moments $\vec{d}_e$ and a magnetic moments $\vec{d}_m$ in an external electromagnetic field $(\vec{E},\vec{B})$, 
$\vec{A}=(\vec{d}_m \times \vec{E}+\vec{d}_e\times\vec{B})/\hbar c$ \cite{Furtado, Pachos, Pachos2, Kay}. For a rotating lattice, $\vec{A}=m\vec{\Omega}\times\vec{r}/\hbar$, where $\vec{\Omega}$ is the rotation frequency and $m$ is the mass of the atom.
In this work, we study the case of the uniform effective magnetic field $\vec{B}=\vec{\nabla}\times\vec{A}=B\hat{z}$. 
Results will depend on the frustration parameter $\alpha$, defined as the flux per plaquette in units of $2\pi$,
\begin{equation}\label{eq:Frustration}
\alpha=\frac{1}{2\pi}\int\vec{B}\cdot d\vec{S}_{\rm plaq} = \frac{1}{2\pi} \sum_{\rm plaq} \phi_{ij}
\end{equation}
where the integration is over the surface of a lattice plaquette and
the sum is performed anticlockwise over the edges of the square
plaquette. This parameter is only meaningful between 0 and 1 because a
flux of $2\pi$ through a plaquette has no effect on the
system. Frustration is maximal at $\alpha=1/2$.

In this paper, we will use a magnetic analogy
as the framework to study the Bose-Hubbard problem.  This is most
easily motivated in the limit of $U/t \to \infty$, even though we will
not be working directly in this limit. In such a limit, the site
occupation can be restricted to zero and one boson. Then, the Hilbert
space of possible states can be mapped onto a spin-half XY model. The
two $S_z$ states of the pseudospin correspond to whether a lattice
contains a boson or not.

The spin raising and lowering operators correspond to the creation and annihilation of
hard-core bosons, respectively.
This mapping is possible because hard-core bosons have the same
commutation relations as $S=1/2$ operators: operators on different
sites commute but operators on the same site anticommute. 
The motion of the atoms translates to pseudospin exchange. The
effective Hamiltonian is
\begin{equation}\label{eq:easyplanemagnet}
H_{eff}= -\frac{J}{2}\sum_{\langle ij\rangle}
\left(e^{i\phi_{ij}}\hat{S}_i^+\hat{S}_j^{-}+\mbox{H.c.}\right)-h\sum_j\hat{S}_j^z
\end{equation}
where $J=2t$,
$\hat{S}_{i}^{\pm}=\hat{S}^x_i \pm \hat{S}^y_i$ are the spin-$1/2$
operators, and $h=\mu$ represents an effective Zeeman field. Note that
this is a ferromagnet in the absence of frustration ($\phi_{ij} = 0$).

It is not simple to attack the infinite-$U$ limit of
the problem of hard-core boson directly.  Instead, we will relax the
hard-core condition and allow for more than one boson on each site. We
will allow $2S$ atoms on each site so that each site has $2S+1$
possible states. This corresponds to a spin-$S$ model with the
Hamiltonian given in Eq. (\ref{eq:easyplanemagnet}).  The relationship
between the original bosons, $\hat{a}$, and this spin-$S$ model is
established \emph{via} the Holstein-Primakoff representation:
\begin{equation}\label{eq:HolsteinPrimakoffa}
\hat{S}_i^{+} = \hat{c}^\dagger_i(2S -\hat{c}^{\dagger}_i\hat{c}_i)^{1/2},
\qquad\hat{S}_i^z=\hat{c}^{\dagger}_i\hat{c}_i - S.
\end{equation}
where $\hat{c}_i$ are operators with bosonic commutations and are essentially
the original bosons $\hat{a}_i$ of the Bose-Hubbard model.
The limit of $S\rightarrow\infty$ corresponds to the classical limit
of the model. More specifically, we need $S\rightarrow\infty$ while
$JS$ and $h$ remain constant so that exchange and Zeeman energies
remain comparable.

Mathematically, the large-$S$ limit provides a systematic way to
control the quantum fluctuations in this problem.  Quantum
fluctuations can be introduced (see later) in a $1/S$ expansion under
the assumption that those effects do not alter significantly the
nature of the ordering obtained for the classical ground states. We
will present results to leading order in $1/S$ (i.e., we do not
set $S=1/2$ afterward). Physically, the
leading-order results in $S$ should be relevant to optical
lattices with many atoms per site on average.  

The relaxation of the maximum site occupancy to $2S$ from a model of hard-core bosons
is not the only way to control correlations in the Bose-Hubbard model
at weak tunneling. A similar methodology is to consider a dense but
weakly interacting limit of the Bose-Hubbard model. With $\bar n$
being the average boson density per site, this limit is given by
$U\rightarrow 0$ and $\bar n\rightarrow\infty$ while $U\bar n$ remains
constant \cite{LeeGunn}. Then, one can develop a theory as an
expansion in $1/\bar n$.  This approach produces results very close to
the $1/S$ expansion considered here.

Note that our Hamiltonian has  local gauge invariance. If we change
the gauge, $\vec{A}\rightarrow \vec{A}+\vec{\nabla}\chi$, then
the Hamiltonian stays unchanged if the boson and spin operators pick up a phase change.
\begin{equation}\label{eq:GaugeTransformation}
\phi_{ij}\rightarrow e^{i(\chi_j-\chi_i)}\phi_{ij}\,, \quad
\hat{a}_i\rightarrow e^{i\chi_i}\hat{a}_i\,,\quad
\hat{S}^-_i\rightarrow e^{i\chi_i}\hat{S}^-_i\,.
\end{equation}
In the spin language, this corresponds to
a rotation of $\chi_i$ in the $xy$ plane in spin space.

Before proceeding to discuss the properties of this system, we point
that we may generalize this to an optical lattice containing two
species of bosonic atoms, such as two hyperfine states.  Let us denote
the two species by $\sigma=\uparrow,\downarrow$.  This allows for more
degrees of freedom in the model Hamiltonian.  Two atomic species may,
in general, see different lattice potentials so that the tunneling
matrix elements and chemical potentials could be different for the two
species.  The Hubbard model for the two species would be of the form
$H_{\rm Hubbard}=H^{(0)}+T$ with
\begin{eqnarray}\label{eq:BoseHubbard2}
H^{(0)}&=&\frac{1}{2}\sum_{i,\sigma,\sigma'}U_{\sigma\sigma'}\hat{a}^{\dagger}_{i\sigma}\hat{a}_{i\sigma'}^{\dagger}\hat{a}_{i\sigma'}\hat{a}_{i\sigma}-\sum_{i,\sigma}\mu_{\sigma}\hat{a}^{\dagger}_{i\sigma}\hat{a}_{i\sigma},\nonumber\\
T&=&-\sum_{\sigma\langle ij\rangle}t_{\sigma}\left(e^{i\phi_{ij}^{\sigma}}\hat{a}_{j\sigma}^{\dagger}\hat{a}_{i\sigma}+\mbox{H.c.}\right),
\end{eqnarray}
where the on-site interaction $U_{\sigma\sigma'}$, the exchange
interaction $t_{\sigma}$, the tunneling phase $\phi_{ij}$, and the
chemical potential $\mu_{\sigma}$ have all acquired a dependence on
the internal states of the bosons. If we specialize to the case of
one atom per site with strong on-site interactions, we can rule out
zero or double occupation of each lattice site. In other words, the
system should be a Mott insulator but the atom occupying each site can
be of either internal state. Thus, each site has a spin-half degree of
freedom: $\hat{S}^+_i = \hat{a}^\dagger_{i\uparrow }\hat{a}_{i\downarrow}$
would create a $\uparrow$ state and $\hat{S}^-_i =
\hat{a}^\dagger_{i\downarrow }\hat{a}_{i\uparrow}$ would create a
$\downarrow$ state.  In this phase, the relative
motion of the two species of atoms is still possible: the motion of one species in one direction must
be accompanied by the motion of the other species in the
opposite direction.  This counterflow keeps the occupation at one atom
at each site. In the pseudospin language, this is simply spin
exchange.  Therefore, in this Mott phase for the overall density, we
have again an easy-plane magnet. If we tune the interactions so that
$U_{\uparrow\uparrow}=U_{\downarrow\downarrow}=2U_{\uparrow\downarrow}$,
then a perturbation theory in $t/U$ brings us to the
effective pseudospin Hamiltonian\cite{Pachos,Kay} described by Eq. (\ref{eq:easyplanemagnet}) with
$J=4t_{\uparrow}t_{\downarrow}/U$,
$h=2\left(\mu_{\uparrow}-\mu_{\downarrow}\right)+8(t_{\uparrow}^2-t_{\downarrow}^2)/U$,
and $\phi_{ij}=\phi_{ij}^{\downarrow}-\phi_{ij}^{\uparrow}$.

We can translate the phases of the single-species Hubbard model to
this two-species system at unit filling.  Superfluidity in the
single-species Hamiltonian at an incommensurate filling corresponds to
superfluidity for counterflow in the two-species problem at the
commensurate filling of one atom per site but with different relative
densities of the two species.  The advantage of considering this
two-species Mott insulator is that there may be more degrees of
freedom in tuning the parameters of pseudospin Hamiltonian, including
the explicit breaking of $S_z \rightarrow -S_z$ spin symmetry.

\section{Classical ground states}
\label{sec:classical}
To determine the ground states of the pseudospin Hamiltonian
(\ref{eq:easyplanemagnet}), we consider first the $S\rightarrow\infty
$ classical ground states for the spin system. We assume that $h>0$
without loss of generality.  In the absence of the vector potential,
the system is an easy-plane ferromagnet.  For $h < h_c =
4JS$, the ground state has a uniform magnetization in the
$xy$ plane in spin space. The $xy$ component of the magnetization at each site
is $m_{xy} = [1-(h/h_c)^2]^{1/2}$.  This $xy$ magnetization corresponds to superfluidity in the original single-species Hubbard model.  The
$z$ magnetization in the $S^z$ direction $M_z = N\langle S^z_i\rangle
= Nh/h_c$ corresponds to the number of atoms in the optical lattice
measured from half filling. For higher Zeeman fields ($h>h_c$), $M_z$
becomes saturated and there is no $xy$ magnetization: the lattice is a
Mott insulator at one atom per site (or empty for $h<-h_c$).

\begin{figure}[htb]
\includegraphics[width=\columnwidth]{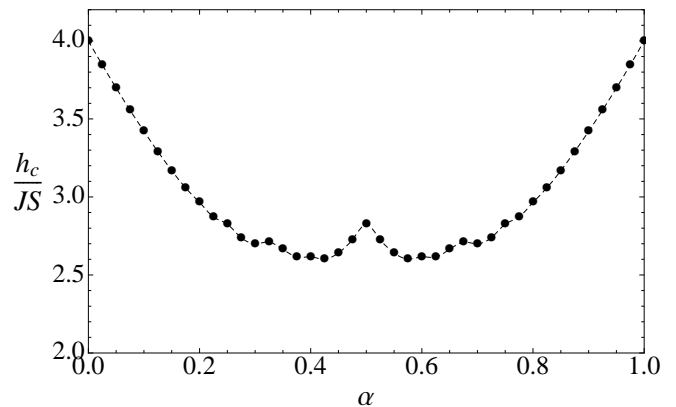}
  \caption{Critical value of the effective Zeeman field, $h_c(\alpha)$, as a
 function of the parameter $\alpha$ being the flux per plaquette in units of $2\pi$. For $h >  h_c(\alpha)$ the lattice is a Mott insulator at one atom per site.}  \label{fig:hofstadter}
\end{figure}

In the presence of the vector potential, the ordering pattern of the
classical ground state depends on the effective magnetic flux through
each plaquette. This introduces vortices into the spin pattern. It
also reduces the critical field $h_c$ below which the
$xy$ magnetization is nonzero. As shown by P\'azm\'andi and Domanski\cite{Pazmandi}, $h_c$ is
given by is the maximal eigenvalue of the matrix
$JSe^{i\phi_{ij}}$. This is shown in
Fig.~\ref{fig:hofstadter}. Note that this result for $h_c$ is not
restricted to the classical limit but applies for all values of the
spin $S$.  The spectrum of all the eigenvalues of this matrix as a
function of the frustration parameter $\alpha$ is the Hofstadter
spectrum \cite{Hofstadter} as discussed originally in terms of two-dimensional
tight-binding electrons in the quantum Hall regime.

Let us now turn to the classical ground states for $h < h_c$. Writing
the local magnetization in spherical polars,
$\langle \vec{S}_i \rangle =
S(\sin\theta_i \cos\phi_i,\sin\theta_i\sin\phi_i,\cos\theta_i)$, the
classical energy is given by:
\begin{equation}\label{eq:ClassicalEnergy}
E^{\rm class} \simeq 
-JS^2\!\!\sum_{\langle ij\rangle}
\sin\theta_i\sin\theta_j\cos\left(\phi_i-\phi_j+\phi_{ij}\right)
-hS\!\!\sum_i\cos\theta_i.
\end{equation}
Minimizing this energy, we find that the ground-state values 
for $\phi_i$ and $\theta_i$, $\Phi_i$ and $\Theta_i$, must satisfy, for each
site $i$, 
\begin{eqnarray}\label{eq:ClassicalEquations}
JS\sin\Theta_i\sum_{j= i+\delta}\sin\Theta_j 
    \sin\left(\Phi_i-\Phi_j+\phi_{ij}\right)&=&0\nonumber\\
JS\cos\Theta_i\sum_{j= i+\delta}\sin\Theta_j 
    \cos\left(\Phi_i-\Phi_j+\phi_{ij}\right)&=&h\sin\Theta_i 
\end{eqnarray}
where the summation is taken over the four neighboring sites of $i$:
$j=i+\delta$.  The first equation conserves the spin current (or
atomic current in the original Hubbard model) at each node.  The
second specifies that there is no net effective Zeeman field causing
precession around the $z$ axis in spin space.  In the original boson
language, this ensures a uniform local chemical potential throughout
the system (in the Hartree approximation).  The system has a local
gauge invariance and we need to fix a gauge to perform our numerical
calculations. We choose the Landau gauge $\vec{A}=B\left(0,x,0\right)$
so that the Aharonov-Bohm phase $\phi_{ij}$ is zero on all horizontal
bonds of the lattice.
\begin{figure}[bht]
 \includegraphics[width=\columnwidth]{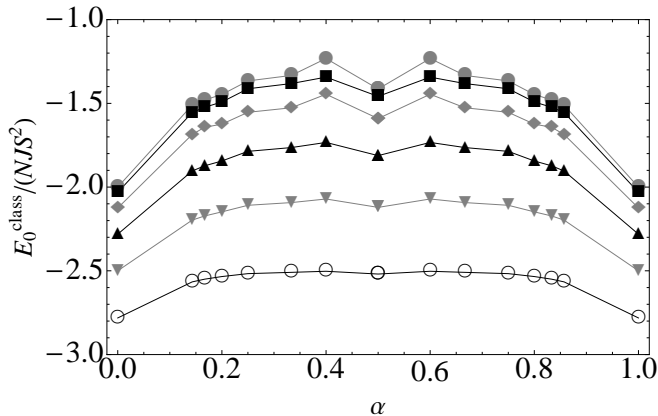} 
 \caption{Ground
 state energy of the classical spin system as a function of the
 frustration parameter $\alpha$ (flux per plaquette divided by $2\pi$)
 for different Zeeman fields $h/JS = 0, 0.5, 1, 1.5, 2$ and 2.5 (from top to bottom). 
 The energy is symmetric around the point $\alpha=1/2$.}
  \label{Fig.2}
\end{figure}
The classical ground states are obtained by using the Metropolis
algorithm. For rational values of the frustration parameter
$\alpha=p/q$, the Monte Carlo simulations are done on $nq\times nq$
lattices with periodic boundary conditions. In most cases, we find
that the periodicity of the ground state is $q\times q$. However, we
also find ground states with the periodicity $2q\times2q$ in some
cases. The ground-state energies as functions of the flux through a
plaquette are shown in Fig.~\ref{Fig.2}. 

\begin{figure}[hbt]
\includegraphics[width=\columnwidth]{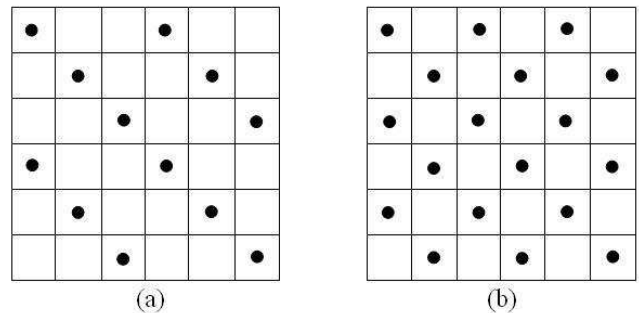}
\caption{\small{Vortex patterns for (a) $\alpha=1/3$ and (b) $\alpha=1/2$ (chequerboard configuration), with $\alpha$ being the flux per plaquette in units of $2\pi$. For $ \alpha=1/3$ there are $2q=6$ degenerate states (vortices can be on three different $3\times 3$ sublattices and along both diagonals). For $\alpha=1/2$ there are two degenerate states with vortices at one or the other diagonal.}}
 \label{Fig.3}
\end{figure}

\begin{figure}[hbt]
\includegraphics[width=\columnwidth]{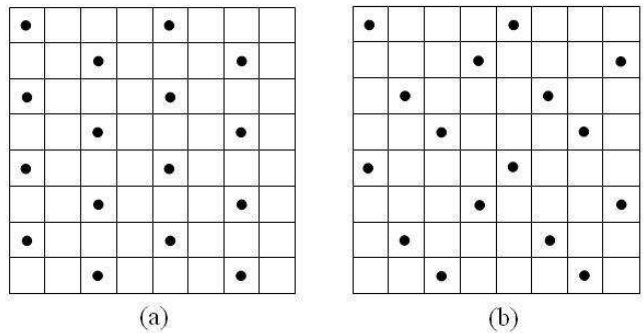}
\caption{\small{Vortex patterns for two ground states at $\alpha=1/4$ and $h=0$. (a) Current pattern periodic on $4\times4$ square, phase pattern periodic on $8\times8$ square. (b) Current and phase patterns periodic on $4\times4$ squares.}}\label{Fig.4}
\end{figure}

We can also examine the vortex pattern in these ground states.
The current on the bond joining sites $i$ and $j$ is given by:
$I_{ij}=(JS^2/\hbar)
\sin\Theta_i\sin\Theta_j\sin\left(\Phi_i-\Phi_j+\phi_{ij}\right)$. The circulation of these currents around each plaquette gives the vortex patterns. These are shown for 
$\alpha=1/2,1/3$, and $1/4$ in Figs. \ref{Fig.3}
and \ref{Fig.4}. 


In case of a zero Zeeman field $h=0$, the classical Hamiltonian
(\ref{eq:ClassicalEnergy}) has been studied extensively in the context of 
Josephson junction arrays in the presence of a perpendicular magnetic field
\cite{Halsey, Straley, Teitel}. 
Halsey \cite{Halsey} showed that, for simple fractions in the range
$1/3\leq\alpha\leq1/2$ (e.g.,  $\alpha=1/2,1/3,2/5,3/7,3/8$),
the ground states have a constant
current along diagonal staircases. Our results for $h=0$ agree with
these previous studies. For a general nonzero Zeeman field, the
ground states we found for $\alpha=1/2$ and $1/3$ also have currents
in diagonal staircases. We cannot obtain analytic generalization of the Halsey solution for the case of finite $h$. We find the ground states by using the Metropolis algorithm. At finite $h$, the phase patterns for $\alpha=1/2$ and $\alpha=1/3$ are similar to the phase patterns for the Halsey states at $h=0$ but $S^z$ has spatial variation around a finite average. 

The Halsey analysis does not cover cases when $\alpha<1/3$.  At
$\alpha=1/4$ and $h=0$ we find two distinct ground state
configurations (Fig.~\ref{Fig.4}) with the same energy in the
agreement with previous results \cite{Straley, Teitel, Kasamatsu}. For
both configurations, the current patterns are periodic on $4\times 4$
square. However, the phase patterns do not have the same periodicity:
it is $8\times 8$ periodic in the configuration shown in
Fig.~\ref{Fig.4} (a) but $4\times 4$ in Fig.~\ref{Fig.4} (b). We find states of the
form [Fig.~\ref{Fig.4} (b)] for general $h$ when simulations are done on $4\times4$
lattices with periodic boundary conditions. 
Simulations done on larger $4n\times4n$ lattices at nonzero $h$
give states that contain elements of both structures separated by
domain walls. Similar results were found by Kasamatsu \cite{Kasamatsu}.

\section{Excitation Spectrum}
\label{sec:quantum}
In this section, we compute the excitations of the system using the spin-wave theory. Quantum effects are incorporated in the problem by
considering finite values of $S$. We will perform an expansion in
powers of the parameter $1/S$ and keep only the terms of the lowest order
in $1/S$ in the Hamiltonian. Even though we are interested in $S\sim$
O(1), the large-$S$ approach is
in some cases justified due to the good convergence of the
perturbative series \cite{Canali,Igarashi,Chubukov,Kollar,Bernardet}. Spin-wave
approximation relies on an assumption that the introduction of the
quantum fluctuations does not qualitatively change the nature of the
ordering obtained for classical ground state. We use this approach to
investigate whether the Bose condensate becomes unstable in any
parameter regime.

Starting from the classical ordered state, we use the
Holstein-Primakoff transformation to represent the spin flips away from the classical ground state in terms of the bosonic operators. We will keep only the
quadratic terms in the final bosonic Hamiltonian. It is convenient to introduce the operators
$\hat{\vec{\mathbb{S}}}_i$ such that $\hat{\mathbb{S}}_i^x$ direction
is parallel to the classical spin direction at each site
\begin{equation}\label{eq:rotatespins}
\left[\begin{array}{c}\hat{\mathbb{S}}_i^x\\\hat{\mathbb{S}}_i^y\\\hat{\mathbb{S}}_i^z\end{array}\right]=\left[\begin{array}{ccc}\sin\Theta_i\cos\Phi_i& \sin\Theta_i\sin\Phi_i&\cos\Theta_i\\-\sin\Phi_i&\cos\Phi_i&0\\ -\cos\Theta_i\cos\Phi_i&-\cos\Theta_i\sin\Phi_i&\sin\Theta_i\\\end{array}\right]\left[\begin{array}{c}\hat{S}_i^x\\\hat{S}_i^y\\\hat{S}_i^z\end{array}\right],
\end{equation}
and use the Holstein-Primakoff representation of these new spin
operators in terms of the bosonic operators, $\hat{b}_i$,
\begin{equation}\label{eq:HolsteinPrimakoffb}
\hat{\mathbb{S}}_i^{+}\equiv \hat{\mathbb{S}}_i^y+ i\hat{\mathbb{S}}_i^z 
= (2S -\hat{b}^{\dagger}_i\hat{b}_i)^{1/2}\hat{b}_i,
\qquad\hat{\mathbb{S}}_i^x=S-\hat{b}^{\dagger}_i\hat{b}_i.
\end{equation}
Note that a gauge transformation corresponds to a rotation of the spin
$\vec{S}$ around the $z$ axis.  Since these new spin variables are
aligned with the classical spin configuration (whatever the choice of
gauge), the new spin $\vec{\mathbb{S}}$ is \emph{invariant} under such
rotation. Therefore, the bosonic operators, $\hat{b}_i$, are gauge invariant.

Under assumption that the zero-point fluctuations are small so that the
average number of spin flips at each site is small compared to $S$, we can
approximate $[1-\hat{b}^\dagger_i \hat{b}_i/(2S)]^{1/2}$ as unity. The
resulting Hamiltonian, to order O($S^0$), is
\begin{equation}\label{eq:HamFluctuations}
\hat{H}\simeq 
E^{\rm class}_0 +\sum_{\langle ij\rangle}
\left(A_{ij}^{-}\hat{b}_i\hat{b}_j-A_{ij}^{+}\hat{b}_i\hat{b}_j^{\dagger}+{\rm H.c.}\right)+
\sum_iC_i \hat{b}_i^{\dagger}\hat{b}_i,
\end{equation}
with
\begin{eqnarray}\label{eq:HamFluctuationsCoeffs}
A_{ij}^{\pm}&=&\frac{JS}{2}\left[(\cos\Theta_i\cos\Theta_j\pm 1)\, c_{ij}
\pm i(\cos\Theta_i\pm\cos\Theta_j)\, s_{ij}\right],\nonumber\\
C_{i}&=&JS\sin\Theta_i\sum_{j=i+\delta}\sin\Theta_j c_{ij}+ h\cos\Theta_i
\end{eqnarray}
where $c_{ij}=\cos(\Phi_i-\Phi_j+\phi_{ij})$,
$s_{ij}=\sin(\Phi_i-\Phi_j+\phi_{ij})$ and $E^{\rm class}_0$ is the
ground-state value of the classical energy [Eq. (\ref{eq:ClassicalEnergy})].
Note that all the coefficients
in this Hamiltonian are gauge invariant, confirming our above conclusion that 
the bosonic operators, $\hat{b}_i$, are gauge invariant. 

This Hamiltonian also reduces correctly to the case of $h>h_c$
(i.e., $\Theta_i=0$) when there is no need for realigning the axis of
quantization [Eq. (\ref{eq:rotatespins})]. In that case, the ``anomalous''
terms $\hat{b}\hat{b}$ and $\hat{b}^\dagger\hat{b}^\dagger$ in the
Hamiltonian vanish. Then, the spin excitations are described by a
tight-binding model with magnetic flux through the plaquettes:
\begin{equation}\label{eq:HamFluctuationsSaturated}
\hat{H}_{h\geq h_c}\simeq -hNS-JS\sum_{\langle ij\rangle}
\left(e^{i\phi_{ij}}\hat{b}_i\hat{b}_j^{\dagger}+{\rm H.c.}\right)+h\sum_i\hat{b}_i^{\dagger}\hat{b}_i.
\end{equation}
This is diagonalized by the Hofstadter
solution \cite{Hofstadter}. The excitation spectrum has an energy gap
of $h-h_c$ and the ground state corresponds to a vacuum of these
excitations, i.e., there are no zero-point fluctuations in the
ground state.

For lower Zeeman fields ($h<h_c$), Hamiltonian
(\ref{eq:HamFluctuations}) containing the anomalous terms will have
zero-point fluctuations which reduce the magnetization from the
classical value. In the language of the original bosons, the
fluctuations would deplete the condensate. The Hamiltonian can be
diagonalized by a generalized Bogoliubov transformation,
\begin{equation}\label{eq:Bogoliubov}
\hat{b}_i = \sum_m \left(u_{im} \hat{\alpha}_m + v^*_{im} \hat{\alpha}^\dagger_m\right)
\;,\quad
\hat{b}^\dagger_i = \sum_m \left( v_{im} \hat{\alpha}_m + u^*_{im} \hat{\alpha}^\dagger_m\right)
\end{equation}
for $m=1,\ldots$, $I$ for a lattice of $I$ sites. To ensure that
the new operators $\hat{\alpha}_m$ obey bosonic commutation relations, we require the matrices
$\mathbf{u}$ and $\mathbf{v}$ to obey:
$\mathbf{u}\mathbf{u}^\dagger - \mathbf{v}\mathbf{v}^\dagger = \mathbf{1}$
and $\mathbf{u}\mathbf{v}^{\rm T} - \mathbf{v}\mathbf{u}^{\rm T} = \mathbf{0}$.
To obtain a diagonalized Hamiltonian in terms of these new operators, we can write 
the part of the Hamiltonian (\ref{eq:HamFluctuations}) quadratic in the bosonic operators 
as $\hat{H} = \hat{c}^\dagger M \hat{c}$, where $M$ is a $2I\times 2I$ matrix and 
$\hat{c} = (\mathbf{b},\mathbf{b}^\dagger)$ with $\mathbf{\hat{b}}=(\hat{b}_1,\hat{b}_2,...)$.
Then, it can be shown that Hamiltonian (\ref{eq:HamFluctuations}) is diagonalized 
into the form
\begin{equation}\label{eq:HamDiagonalised}
 \hat{H} = E_0 + \sum_m \epsilon_m \hat{\alpha}^\dagger_m \hat{\alpha}_m 
\end{equation}
with eigenenergies $\epsilon_m$
if we solve the eigenvalue problem,
\begin{equation}\label{eq:Mequation}
\left(M-\frac{\epsilon}{2}\Sigma_z\right)q=0.
\end{equation}
where $ q_m = (u_{1m},\ldots,u_{Nm},v^*_{1m},\ldots,v^*_{Nm})$ contains the coefficients
of the Bogoliubov transformation
and $\Sigma_z=\left\{\left\{\mathbf{1},\mathbf{0}\right\},\left\{\mathbf{0},\mathbf{-1}\right\}\right\}$. 

\begin{figure}[hbt]
 \includegraphics[width=\columnwidth]{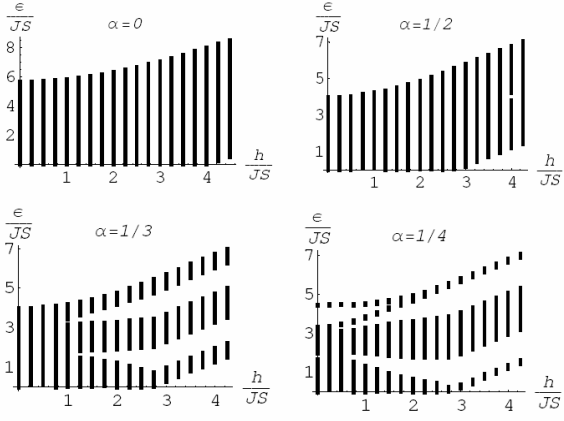}
 \caption{Low energy excitation spectrum as a function of
  the Zeeman field $h$ for $60\times60$ lattices with periodic
  boundary conditions for frustration $\alpha=0,1/4,1/3$ and $1/2$. Critical values $h_c$ are: $h_c(\alpha=0)=4$,
  $h_c(\alpha=1/4)= 2.828$, $h_c(\alpha=1/3)= 2.732$ and
  $h_c(\alpha=1/2)= 2.828$. Above $h_c$, the spectrum has a finite energy
  gap. The spectrum is gapless for $h<h_c$ indicating long-range order in the
  system.}  \label{Fig.5}
\end{figure}

We computed the spectrum for $60\times60$, $120\times120$ and
$240\times240$ lattices with periodic boundary conditions, using the
classical ground states from our Monte Carlo simulations discussed in
the previous section.  Our results for $60\times 60$ lattices and the
frustration parameters $\alpha =$ 0, 1/2, 1/3, and 1/4 are shown in
Fig.~\ref{Fig.5}. Our result for $\alpha=1/4$ is calculated using the
$4\times4$ periodic classical ground state presented in
Fig.~\ref{Fig.4}(b).

As can be seen in Fig. \ref{Fig.5} at $h<h_c(\alpha)$, the spectrum
is gapless. The low-energy excitations are the Goldstone modes related
to the spontaneous symmetry breaking of the global rotation symmetry in
the $xy$-plane in spin space. In other words, the spin system has
long-range magnetization in the $xy$ plane in spin space. We can use
$\langle S^+_i \rangle$ as the order parameter. In the language of the
original bosonic model, this corresponds the breaking of U(1) symmetry
due to Bose condensation. Above $h_c$, there is no symmetry breaking
and we see an energy gap in the system proportional to $h-h_c$ as
discussed above.

\begin{figure}[htb]
\includegraphics[width=\columnwidth]{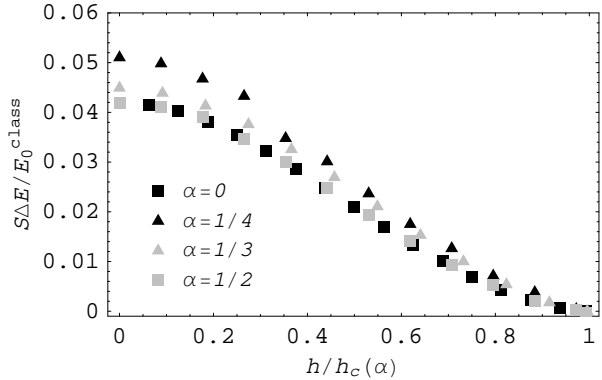}
\caption{Quantum correction to the ground-state energy as a function
  of $h/h_c(\alpha)$ for $\alpha = 0,1/2,1/3$ and $1/4$. $h_c(\alpha)$
  is the critical value of the Zeeman field $h$ for a given
  frustration parameter $\alpha$.}
 \label{Fig.6}
\end{figure}

The ground-state energy $E_0$ [Eq. (\ref{eq:HamDiagonalised})] can be
written as $E^{\rm class}_0+\Delta E_0$, where $\Delta
E_0=\Delta+\sum_m \epsilon_m/2$ is a quantum correction to the
classical ground-state energy [Eq. (\ref{eq:ClassicalEnergy})] with
$\Delta=-JS\sum_{\langle ij\rangle}
\cos(\Phi_i-\Phi_j+\phi_{ij})$ for $h=0$ and
$-h\sum_i1/(2\cos\Theta_i)$ for $h\neq 0$.  This quantum correction is
of order $S^0$ while the classical energy is of order $S$ and so the
fractional change is small in the large-$S$ limit.  We calculate the
relative corrections $\Delta E_0/E^{\rm class}_0$ for several lattice
sizes ($60\times60$, $120\times120$, $240\times240$) and extrapolate
results to the thermodynamic limit shown in Fig. \ref{Fig.6}.  As
can be seen, the quantum correction decreases to zero as the Zeeman
field $h$ approaches the critical value $h_c$.  Above $h_c$, the
ground state is the classical ground state containing no zero-point
fluctuations.

\section{Density Matrix}
\label{sec:densitymatrix}
In this section, we will examine ODLRO in the density matrix \cite{Yang, Penrose}. Consider first the case
without a vector potential. A macroscopically large
eigenvalue of the density matrix $\rho_{ji}$ signals the existence of
Bose-Einstein condensation for our boson problem. Since we are
considering a lattice system above half filling, it is more meaningful
to consider the condensation of vacancies because this is the most
appropriate description as $h$ approaches $h_c$. (For the two-species
model with counterflow superfluidity, we are considering the
condensation of the minority species.)  The hole density matrix is
defined as $\rho^{\rm h}_{ji} = \langle a_i a^\dagger_j\rangle$. The
existence of a macroscopic eigenvalue, $N_0$, corresponds to
Bose-Einstein condensation. The sum of all non-macroscopic eigenvalues
gives the number of holes not in condensate and we can define the
fractional condensate depletion as the ratio of the non-macroscopic
sum to the total number of holes $N_{\rm h}$ which is the trace of the
density matrix.

In the analog of the easy-plane magnet, we should study the
spin-spin correlation function for the spin components in the
$xy$ plane: $\rho_{ji}=\langle\hat{S}_i^{-}\hat{S}_j^+\rangle$.  ODLRO
corresponds to a non-zero $xy$ magnetization which is the analog of
Bose condensation.  In the large-$S$ limit, $\rho_{ji}/2S$ is the
analog of the bosonic hole density matrix $\rho^{\rm h}_{ji}$ for
$h$ close to $h_c$.

The macroscopic eigenvalue for our spin-spin correlation function is,
to the leading order in $S$, given by the classical value $N_0^{\rm
  class}=\sum_i(m^{xy}_i)^2$, where $\vec{m}^{xy}_i$ is the classical
value of the magnetization at site $i$. We present below our results
for condensate and the depletion of the condensate, i.e., zero-point fluctuations which
decrease the magnetization in the ground state.

The above discussion needs to be modified in the presence of a vector
potential because the density matrices, $\rho$ and $\rho^{\rm h}$, are
not gauge-invariant quantities: $\rho_{ji}\rightarrow
e^{i(\chi_i-\chi_j)}\rho_{ji}$ under the gauge transformation [Eq. (\ref{eq:GaugeTransformation})].  However, we can construct
gauge-invariant analogs. Moreover, the eigenvalues of $\rho$ and
$\rho^{\rm h}$ are gauge invariant even though the corresponding
eigenvectors are not.  Consider first the spin-spin correlation
function in the ground state
\begin{eqnarray}\label{eq:DensityMatrix1}
\rho_{ji}&=&\langle\hat{S}_i^{-}\hat{S}_j^+\rangle=\rho_{ji}^{\rm class}+
\delta\rho_{ji}\nonumber\\
\rho_{ji}^{\rm class} &=& \psi^*_i \psi_j\quad\mbox{with}\quad
\psi_i=Se^{i\Phi_i}\sin\Theta_i
\end{eqnarray}
where $\rho_{ji}^{\rm class}$ is the classical value of the density
matrix (of order $S^2$) and $\psi_i$ is the classical value of the
order parameter (of order $S$) $\langle\hat{S}_i^+\rangle$.  The order
parameter itself is reduced by quantum fluctuations,
\begin{equation}\label{eq:OrderParameter}
\langle\hat{S}_i^+\rangle= \psi_i (1 - \Delta_i)
\,,\quad \Delta_i = \frac{1}{S}\sum_m|v_{im}|^2.
\end{equation}
The correction $\delta\rho$ to the density  matrix is given by:
\begin{equation}\label{eq:deltarho}
\delta\rho_{ji} \simeq - \rho^{\rm class}_{ji} (\Delta_i + \Delta_j) + 
\frac{S}{2} e^{i(\Phi_j-\Phi_i)}\sum_n q_{jn}^* q_{in}
\end{equation}
where $q_{in}=u_{in}+v_{in}+\cos\Theta_i(v_{in}-u_{in})$, with
$u_{in}$ and $v_{in}$ being the coefficients for the Bogoliubov
transformation [Eq. ($\ref{eq:Bogoliubov}$)]. This density matrix is not
invariant under a gauge transformation. We obtain a gauge-invariant
version of the density matrix by expressing it with respect to a
gauge-covariant basis. The most natural basis is the basis formed by
the eigenvectors of the classical density matrix $\rho^{\rm class}$.
The eigenvector corresponding to the largest eigenvalue is simply
$\psi_i$,
\begin{eqnarray}\label{eq:DensityMatrixClassical}
\sum_i\rho_{ji}^{\rm class}\psi_i&=&N_0^{\rm class}\psi_j
\quad\mbox{with}\nonumber\\ N_{0}^{\rm class}=\sum_i
|\psi^*_i\psi_i|^2&=&S^2\sum_i\sin^2\Theta_i
\end{eqnarray}
where $N_0^{\rm class}$ is simply the classical value of the sum of the square of the
$xy$ magnetization ($m_{xy}^2$) on each site.  It is on the
order of $NS^2$ at $h=0$ and tends to zero as $h$ reaches $h_c$.  All
the other eigevectors of $\rho^{\rm class}$ have eigenvalues of zero.
Using an orthonormal set of these eigenvectors as columns for a
unitary matrix $U$, we can construct a unitary transformation for the
density matrix ($\rho\rightarrow\tilde\rho$, etc.),
\begin{equation}\label{eq:DensityMatrixInvariant}
\tilde{\rho}=U^{\dagger}\rho U=\tilde{\rho}^{\rm class}+\delta\tilde{\rho}\,.
\end{equation}
where $\tilde\rho^{\rm class} = \mbox{diag}(N_0^{\rm
  class},0,\ldots,0)$.  Under the gauge transformation
[Eq. (\ref{eq:GaugeTransformation})], all the eigenvectors of $\rho_{ji}$
pick up a phase change, e.g., $\psi_i \rightarrow
e^{-i\chi_i}\psi_i$ so that $U_{ij}\rightarrow e^{-i\chi_i}U_{ij}$. It
is easy to check that this compensates for the phase change in
$\rho_{ji}$ so that $\tilde{\rho}_{ij}\rightarrow\tilde{\rho}_{ij}$.
Consequently, all the quantities obtained from the matrix
$\tilde{\rho}$ are gauge-invariant and therefore physically
meaningful.  In this section, we calculate the effect of quantum
fluctuations on the density matrix.  This requires only the
eigenvalues of $\tilde{\rho}$. They are in fact the \emph{same} as the
eigenvalues of $\rho$ because the two density matrices are related by
a unitary transformation.

\begin{figure}[htb]
\includegraphics[width=\columnwidth]{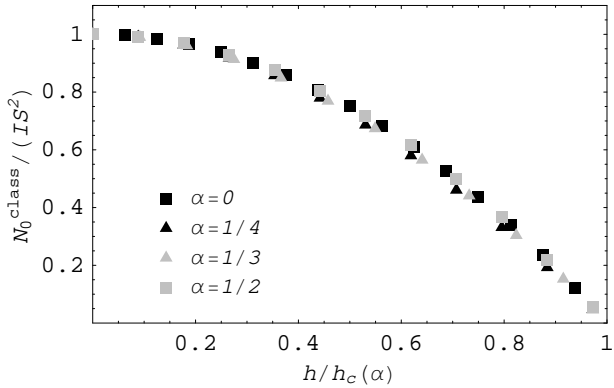}
\caption{The condensate density per site, $N_0^{\rm classical}/I$, in the classical limit for $\alpha=0,1/4,1/3,1/2$. ($I=$ number of lattice sites.) $h_c(\alpha)$ is the
  critical value of the Zeeman field $h$ for a given frustration
  parameter $\alpha$.}
 \label{fig:classicalN0}
\end{figure}

We will now present our numerical results.
First of all, we present the classical solution for the number of atoms in the condensate, $N_0^{\rm class}$, as given by Eq. (\ref{eq:DensityMatrixClassical}).  This is shown in Fig.~\ref{fig:classicalN0}. We see that this decreases to zero as $h$ is increased to $h_c(\alpha)$.

Next, we compute the quantum corrections to the classical solution. In the large-$S$ expansion, these corrections are small and the leading corrections are of order $1/S$ compared to the classical limit. We have computed this leading-order correction and present results in terms of the correction to the classical limits as \emph{fractions} of the classical solution.

\begin{figure}[hbt]
\includegraphics[width=\columnwidth]{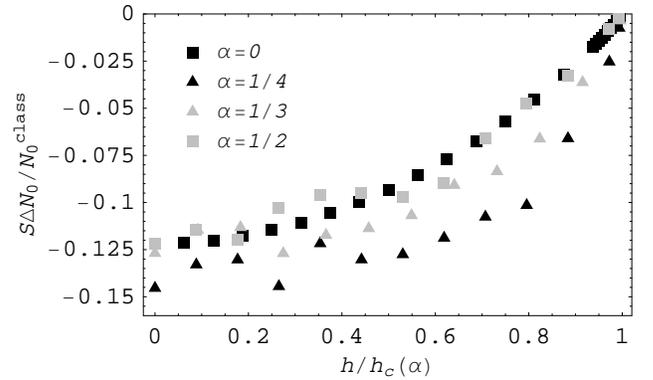}
\caption{Quantum correction $\Delta N_0$ to the the macroscopic
  eigenvalue of the density matrix as a function of $h/h_c(\alpha)$
  for $\alpha=0,1/4,1/3$ and $1/2$. Results have been extrapolated to
  the thermodynamic limit ($L\rightarrow\infty$). $h_c(\alpha)$ is the
  critical value of the Zeeman field $h$ for a given frustration
  parameter $\alpha$.}
\label{Fig.7}
\end{figure}

We can exploit the large-$S$ expansion to compute the eigenvalues of the
density matrix. 
We start with calculating the quantum correction to
the non-degenerate macroscopic eigenvalue, $N_0$.
Since $\rho_{ji}^{\rm class}$ is larger than
$\delta\rho_{ji}$ by an order in $S$, we can calculate the
eigenvalues of $\rho$ by treating $\delta\rho$ in
perturbation theory. The first-order correction to $N_0$ is then given by
\begin{equation}\label{eq:N0Correction}
\Delta N_0=\frac{1}{N_0^{\rm class}}\sum_{ij}
\psi_i^{*}\,\delta\rho_{ij}\,\psi_j = \delta\tilde\rho_{11}
\end{equation}
if the first basis vector for $\delta\tilde\rho$ is chosen to be the
one corresponding to the classical solution $\psi$.  This correction
is of order $S$, as opposed to order $S^2$ for the classical value.
Our results for $\Delta N_0$ as a fraction of $N_0^{\rm class}$ are
shown in Fig.~\ref{Fig.7}. We see that the reduction in $N_0$ is
largest at $h=0$ and decreases to zero at the critical fields
$h_c(\alpha)$.  The vanishing of quantum corrections as $h\rightarrow
h_c$ ($\Theta_i\rightarrow 0$) can be seen directly from the
coefficients $A^-$ of the anomalous terms in Hamiltonian
(\ref{eq:HamFluctuations}) which are responsible for the zero-point
fluctuations in the ground state.

\begin{figure}[hbt]
\includegraphics[width=\columnwidth]{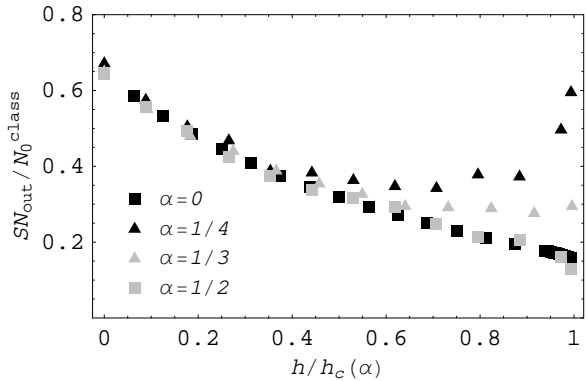}
\caption{Fractional depletion $N_{\rm out}/N^{\rm class}_0$
  for $\alpha=0,1/4,1/3$ and $1/2$ and as a function of
  $h/h_c(\alpha)$. $h_c(\alpha)$ is the critical value of the Zeeman
  field $h$ for a given frustration parameter $\alpha$. Results have
  been extrapolated to the thermodynamic limit
  ($L\rightarrow\infty$).}
\label{Fig.8}
\end{figure}

We can also calculate the sum of the
non-macroscopic eigenvalues, $N_{\rm out}$. 
This corresponds to the condensate depletion in the original boson problem.
In the
$S\rightarrow\infty$ limit for a lattice with $I$ sites, the $I-1$ non-macroscopic eigenvalues are all zero.  
The first-order quantum corrections can be obtained using degenerate
perturbation theory --- we can obtain the eigenvalues as the
eigenvalues of the $(I-1)$-dimensional submatrix
$\delta\tilde\rho_{ji}$ for $i,j=2,\ldots,I$ which excludes the
macroscopically occupied state. The sum of these eigenvalues is simply
the trace of the submatrix:
\begin{equation}\label{eq:Nout}
N_{\rm out}=\sum_{i\neq 1}\delta{\tilde{\rho}}_{ii},
\end{equation}
Again, $N_{\rm out}\propto S$ is one order smaller in $S$ than $N^{\rm class}_0$.  
We find that, just as classical condensate density ($N_0^{\rm class}/I$) vanishes as $h\to h_c(\alpha$), 
the out-of-condensate number, $N_{\rm out}$, also vanishes as $h\to h_c(\alpha)$. However, the ratio of the two quantities remains finite. This ratio, $N_{\rm out}/N^{\rm class}_0$, is the \emph{fractional depletion} of the condensate.
This quantity is one of interest in experiments which measure the degree of Bose-Einstein condensation by observing the time of flight of expanding condensates. 
Our results for this fractional depletion $N_{\rm out}/N^{\rm class}_0$, rescaled by $S$,  are shown 
in Fig.~\ref{Fig.8}. 
The occupation of these non-macroscopic modes is also due to the
anomalous terms in the Hamiltonian.  This again should vanish as
$h\rightarrow h_c$. However, Fig.~\ref{Fig.8} shows that the
occupation remains a \emph{finite} fraction of $N^{\rm class}_0$ even
at the critical field $h_c$. In terms of the original boson model,
this result suggests that condensate depletion remains a finite
fraction of the total number of holes even as the hole density
decreases to zero at $h_c$. Our results at zero frustration agrees with previous work\cite{Bernardet,Hen}.

We observe that this fractional depletion decreases monotically as we increase the Zeeman field
$h$ from zero to $h_c$ for $\alpha=0$ and 1/2. For $\alpha=1/3$, the fractional depletion
appears to have zero slope as a function of $h$ near $h_c$. 
Interestingly, for $\alpha=1/4$, the
relative depletion becomes a non-monotonic function of the Zeeman field 
--- the fractional depletion \emph{increases} when $h_c$ is approached.
In fact, if we formally set $S=1/2$, the condensate depletion even reaches
unity before $h$ reaches $h_c$. As we will see in the next section, this change in
behavior for $\alpha=1/4$ is also seen in the superfluid fraction.
We discuss this further in our concluding remarks.

We note that $N_{\rm out} \neq - \Delta N_0$. In other words, the
trace of the density matrix changes due to quantum fluctuations. This
means that, in the quantum magnet, there is more than one possible
measure of ``condensation'' in the ground state. The discrepancy can be
traced to the quantum fluctuations for $S^z$ at each site: $\mbox{Tr
}\rho = \sum_i \langle \hat{S}^+_i \hat{S}^-_i\rangle = \sum_i
[S(S+1)- \langle (\hat{S}^z_i)^2 \rangle + \langle
  \hat{S}^z_i\rangle]$.  For $S=1/2$, this is simply $\sum_i (1/2 +
\langle \hat{S}^z_i\rangle)$, corresponding to the total boson number
in the original model which is a conserved quantity.  However, for any
$S>1/2$, the mean-square fluctuation in the local $z$-component will
alter the total trace of the density matrix.  In other words, this is
an artifact of our large-$S$ generalization of the model.  In the
above, we have compared $N_{\rm out}$ with the macroscopic eigenvalue
$N_0 \simeq N^{\rm class}_0$.  Strictly speaking, in order to discuss
the depletion of the hole condensate in the original boson model, we
should use the analogue for the hole density matrix and then divide
the number of holes in the system.  As discussed above, the
correspondence is simple near $h_c$: we should consider $N_{\rm
  out}/2S$ compared to $\sum_i (S-\langle \hat{S}^z_i\rangle) = S
\sum_i (1-\cos\Theta_i)$. This is qualitatively similar to the results
plotted in Fig.~\ref{Fig.8}.

\section{Superfluid density}
\label{sec:superfluiddensity}
Bose-Einstein condensation can be defined in equilibrium. On the other
hand, superfluidity is related to the transport properties of the
system. Those two phenomena are related through the phase of the
macroscopic wave function (order parameter). The superflow occurs when
the phase of the wave function varies in space. In this section, we
calculate the superfluid density for our system as a response to an
external phase twist. The superfluid density, a characteristic
quantity that describes the superfluid, measures the phase stiffness
under an imposed phase variation and differs from zero only in the
presence of the phase ordering. We find the superfluid fraction
following the calculations of Roth and Burnett\cite{Roth} and Rey \emph{et al.}\cite{Rey} 
where the superfluid density is calculated for the Bose-Hubbard model with real
couplings. Our results show that the superfluid fraction is reduced in
the presence of the frustration.

The superfluid density introduced by considering a change in the free
energy of the system under imposed phase variations \cite{Fisher,
  Roth, Rey} is equivalent to the helicity modulus \cite{Fisher} which
differs from zero only for ordered-phase configurations and is
consequently an indicator of the long-range phase coherence of the
system. The definition is also equivalent to the definition of the
superfluid density in terms of the winding numbers which is used in
the path-integral Monte Carlo methods \cite{Pollock, Paramekanti,
  Scalettar} and to Drude weight or charge stiffness which describes
d.c.~conductivity \cite{Poilblanc, Kohn, Scalpino, Denteneer, Shastry}.

Let us consider a system of size $L_x$ in the $x$ direction. One way
to achieve the phase twist is to impose the twisted boundary
conditions on the wave function describing the system. If we assume
that the phase twist is imposed along the $x$ direction the twisted
boundary conditions are
\begin{equation}\label{eq:SF1}
\Psi^{\bar{\Phi}}\left(\vec{r}_1,...,\vec{r}_i+L_x\hat{x},...\right)=e^{i\bar{\Phi}}\Psi^{\bar{\Phi}}\left(\vec{r}_1,...,\vec{r}_i,...\right)
\end{equation} 
with respect to all coordinates of the wave function. Let us introduce a unitary transformation
\begin{equation}\label{eq:SF2}
U_{\bar{\Phi}}=e^{\sum_i i\chi(\vec{r}_i)}\mbox{ }\mbox{ with }\mbox{ }\bar{\Phi}=\chi\left(\vec{r}+L_x\hat{x}\right)-\chi\left(\vec{r}\right).
\end{equation} 
The untwisted wave function which satisfies the periodic boundary conditions $\Psi(\vec{r}_1,...,\vec{r}_i+L_x\hat{x},...)=\Psi(\vec{r}_1,...,\vec{r}_i,...)$ is related to the twisted wave function via the unitary transformation $U_{\bar{\Phi}}$ as $|\Psi^{\bar{\Phi}}\rangle=U_{\bar{\Phi}}|\Psi\rangle$. The Schr\"odinger equation for the system with twisted boundary conditions, $\hat{H}|\Psi^{\bar{\Phi}}\rangle=E^{\bar{\Phi}}|\Psi^{\bar{\Phi}}\rangle$, can then be rewritten as $\hat{H}_{\bar{\Psi}}|\Psi\rangle=E^{\bar{\Phi}}|\Psi\rangle$ where the twisted Hamiltonian is 
\begin{equation}\label{eq:SF3}
\hat{H}_{\bar{\Phi}}=U_{\bar{\Phi}}^{\dagger}\hat{H}U_{\bar{\Phi}}.
\end{equation}
In other words, the eigenvalues of the twisted Hamiltonian with periodic boundary conditions are the same as eigenvalues of the original Hamiltonian with twisted boundary conditions. 

\begin{figure}[bht]
\includegraphics[width=\columnwidth]{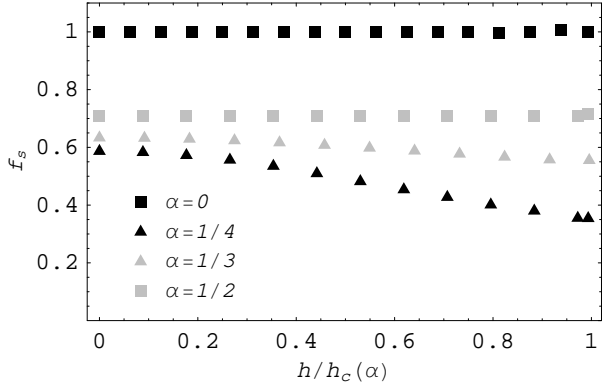}
\caption{Superfluid density as a fraction of the classical condensate density $N_0^{\rm class}/I$ 
as a function $h/h_c(\alpha)$ 
for the frustration parameter $\alpha=0,1/4,1/3$ and $1/2$. $h_c(\alpha)$ is the critical value of the Zeeman field $h$ for a given frustration parameter $\alpha$.}
\label{Fig.9}
\end{figure}
The superfluid velocity is proportional to the order-parameter phase
gradient and an additional phase variation $\chi(\vec{r})$ will change
the superfluid velocity by $\Delta
\vec{v}_s=\hbar\vec{\nabla}\chi(\vec{r})/m$ in the continuous
system. When the imposed phase gradient is small so that other
excitations except increase in the velocity of the superflow can be
neglected the change in the ground-state energy can be approximated by
$\Delta E_g=-\vec{P}\cdot\Delta\vec{v}_s+M_s(\Delta \vec{v}_s)^2/2$,
with $M_s=mN_s$ being the total mass of the superfluid part of the
system. Here we choose a linear phase variation along the $\hat{x}$
direction, $\chi(\vec{r})=\bar{\Phi}x/L_x$. Replacing $\hbar^2/2m$
for the continuous system by $J/2$ for our 2D discrete
lattice we obtain the following expression for the superfluid density
\cite{Roth}
\begin{equation}\label{eq:SF4}
n_s=\frac{I_x}{I_yJ}\frac{\partial^2E_g(\bar{\Phi})}{\partial\bar{\Phi}^2}\left|_{\bar{\Phi}=0}\right.,
\end{equation}
where $I_{x,y}=L_{x,y}/a$ with $a$ being the lattice spacing. The
twisted Hamiltonian is of the same form as the untwisted one only with
$\phi_{ij}$ replaced by $\phi_{ij}-\bar{\Phi}$. Under assumption that
the phase twist $\bar{\Phi}\ll\pi$ we can calculate the ground state
energy of the twisted Hamiltonian perturbatively. Expanding
$e^{i\bar{\Phi}/L_x}$ up to the second order in $\bar{\Phi}$ the
twisted spin Hamiltonian becomes
\begin{equation}\label{eq:SF5}
H^{\bar{\Phi}}=H+\frac{\bar{\Phi}}{I_x}\hat{J}_x-\frac{\bar{\Phi}^2}{2I_x^2}\hat{T}_x,
\end{equation}
where
$\hat{J}_x=iJ\sum_i(e^{i\phi_{ii+x}}\hat{S}_i^+\hat{S}_{i+x}^-
-{\rm H.c.})/2$ is the paramagnetic current operator and
$\hat{T}_x=-J\sum_i(e^{i\phi_{ii+x}}\hat{S}_i^+\hat{S}_{i+x}^-+H.c.)/2$
corresponds to the kinetic-energy operator for the hopping in the
$x$ direction. The terms in the Hamiltonian above that contain the
twist angle can be treated as a small perturbation
$V^{\bar{\Phi}}=\bar{\Phi}\hat{J}_x/I_x-\bar{\Phi}^2\hat{T}_x/2I_x^2$. Calculating
the ground-state energy for the system with imposed small twist within
the second order perturbation theory and using Eq. (\ref{eq:SF4}), we
obtain the following expression for the superfluid density as a fraction
of the condensate density, $f_s=I_xI_yn_s/N_0$:
\begin{equation}\label{eq:SF6}
f_s=-\frac{1}{N_0J}
\left(\langle\psi_0|\hat{T}_x|\psi_0\rangle+
2\sum_{\nu\neq0}\frac{|\langle\psi_{\nu}|\hat{J}_x|\psi_0\rangle|^2}{E_{\nu}-E_0}\right)
\mbox{ , }\mbox{ }\bar{\Phi}\ll\pi,
\end{equation}
where $N_0\simeq N_0^{\rm class}$ in the large-$S$ limit and $\psi_{\nu}$ are
eigenstates of original untwisted Hamiltonian with $\nu=0$ labeling
the ground state. In terms of the original boson model,
$N_0$ corresponds to the number of condensed particles or holes (for $h<0$ or $h>0$). The first term corresponds to the
diamagnetic response of the condensate while the second term
corresponds to the paramagnetic response involving excited states.

The results obtained for the superfluid fraction within the Bogoliubov
approximation are shown in Fig.~\ref{Fig.9}. The leading term due to quantum effects
comes from the paramagnetic term in Eq. (\ref{eq:SF6}). This is
of order $S^0$. In the absence of frustration ($\alpha=0$), the system
is homogenous and the system conserves momentum. This means that the
eigenstates are Bloch states corresponding to different momenta. As a
result, the current matrix element in Eq. (\ref{eq:SF6}), which cannot
couple different momenta, vanishes.  Moreover, the kinetic energy in
the ground state is in itself proportional to $N_0$. In the boson
model, this means that the superfluid fraction corresponds simply to
the kinetic energy per hole.  This is a quantity which is independent
of $h$ and so the superfluid density is the same as the condensate
density in the large-$S$ limit at zero frustration. (However, $1/S$ corrections will change the
result, giving a superfluid density larger than the condensate density
for general $h$, but $f_s \to 1$ as $h\to h_c$.)  Similarly, the
current matrix element vanishes for the fully frustrated case
($\alpha=1/2$). In this case, frustration reduces the superfluid
fraction in $\alpha=1/2$ case to around $70\%$.  For $\alpha=1/3$ and
$1/4$, an increase in the Zeeman field $h$ 
results in a larger reduction in the fraction $f_s$ at values
of $h$ closer to $h_c(\alpha)$. That can be seen in Fig.~\ref{Fig.9}
for the inhomogeneous cases of $\alpha=1/3$ and $1/4$. As for the
condensate depletion, we note that the superfluid density as a fraction of the condensate density does not
vanish as $h\rightarrow h_c$.

We also note that the superfluid density behaves differently for $\alpha = 1/3$
and $1/4$ compared to $\alpha=0$ and $1/2$. The same qualitative change in behavior 
was observed for the condensate depletion calculated in Sec. \ref{sec:densitymatrix}.

\section{Conclusion}
\label{sec:conclusion}
We have studied the ground state for bosonic atoms in a frustrated
optical lattice by mapping the problem to a frustrated easy-plane
magnet. Using a large-$S$ approach, we further introduce quantum
effects under the assumption that those effects do not change
qualitatively the nature of the ordering obtained for the classical
ground states.  We examined our results for any precursor to the
non-superfluid or uncondensed states.

We have found that frustration can decrease the depletion of the
condensate and the superfluid fraction.  However, the fractional
depletion of the condensate and the superfluid fraction remain finite
for all incommensurate filling [$h < h_c(\alpha)$].  The behavior of
the fractional condensate depletion and superfluid fraction as a function of
filling has interesting behavior. We find that the cases of
$\alpha=0$ and 1/2 behave differently from the cases of $\alpha=1/3$
and 1/4.  Surprisingly, for the cases of smaller $\alpha$, the
fractional condensate depletion becomes a nonmonotonic function of
the filling, decreasing as we increase $h$ from zero but eventually
\emph{increases} as $h\rightarrow h_c$.  In fact, if we formally set
$S=1/2$, then the computed fractional depletion exceeds 100\% for the
$\alpha=1/4$ case as $h$ approaches $h_c$.  We also have some evidence
that the same behavior occurs in the $\alpha=1/6$ case for small
system sizes. In other words, our results raise the possibility, for
$\alpha <1/4$, of a second-order phase transition to a non-condensed
state where quantum fluctuations are large enough to destroy Bose
condensation.  It is intriguing to note that this case does not have a
Halsey-type classical ground state and in fact has two degenerate
ground states with different phase patterns. One can speculate that
the motion of domain walls between the two different phase patterns
may contribute to a route to decondensation and/or loss of
superfluidity.

Finally, we note that fractional quantum Hall states are expected when
the number of vortices becomes comparable to the number of atoms or
holes in the Bose-Hubbard model. In our large-$S$ theory, the boson
number is proportional to $S$ and so the quantum Hall regime, if it
exists in such a theory, exists only when $h-h_c \sim 1/S$.
Therefore, one might expect the condensate depletion or the reduction
in the superfluid fraction to be large as $h\rightarrow h_c$.  We do
not find this directly in our perturbative theory in $1/S$. However,
our results for the fluctuations around non-Halsey-type ground states
suggest that an instability to a non-condensed state may be possible.

\end{document}